\documentclass[a4paper,11pt]{article}

\usepackage{amsfonts}
\usepackage{amsmath}
\usepackage{amsthm}
\usepackage[auth-sc,affil-sl]{authblk}
\usepackage{bm}
\usepackage{graphicx} 
\usepackage{caption}
\usepackage{subcaption}
\usepackage{multirow}
\usepackage{verbatim}
\usepackage{enumitem}
\usepackage{longtable}
\usepackage{float}
\usepackage{hyperref}
\hypersetup{
    colorlinks=true,
    linkcolor=blue,
    citecolor=blue,
    urlcolor=black
}
\usepackage{geometry}
\usepackage{underscore}
\usepackage{amssymb}
\usepackage{booktabs}
\usepackage{tabularx}
\usepackage{array}
\usepackage{makecell}
\usepackage{xcolor}
\usepackage{natbib}
\usepackage{siunitx}
\usepackage{pifont}
\newcommand{\cmark}{\ding{51}}   
\renewcommand
\baselinestretch{1.0} 
\usepackage{url}
\usepackage{natbib}
\usepackage{graphicx}
\usepackage{epstopdf} 
\graphicspath{{figures/}} 
\usepackage{lscape}
\usepackage[svgnames]{xcolor}
\usepackage{sansmath}
\usepackage{subcaption}
\usepackage{rotating}
\usepackage{tikz}
\usetikzlibrary{positioning, arrows.meta, shapes.geometric}
\usepackage{xspace}

\DeclareMathAlphabet{\mathpzc}{OT1}{pzc}{m}{it}

\newcommand \address[1]{\gdef \@address{#1}}
\makeatletter
\long\def\@footnotetext#1{\insert\footins{\def\baselinestretch{1.2}\footnotesize
\interlinepenalty\interfootnotelinepenalty
\splittopskip\footnotesep \splitmaxdepth \dp\strutbox
\floatingpenalty \@MM \hsize\columnwidth \@parboxrestore
\edef\@currentlabel{\csname
p@footnote\endcsname\@thefnmark}\@makefntext
{\rule{\z@}{\footnotesep}\ignorespaces #1\strut}}}

\long\def\symbolfootnote[#1]#2{\begingroup%
\def\thefootnote{\fnsymbol{footnote}}\footnote[#1]{#2}\endgroup}

\providecommand{\keywords}[1]{\textit{Keywords:} #1}

\def\maketitle{%
  \null
  \thispagestyle{empty}%
  \begin{center}\leavevmode
    \normalfont
    {\LARGE \bf \@title\par}%
    {\normalsize \@author\par}%
    \vskip 0.05 cm
    \vskip 0.05cm
    {\normalsize \@date\par}%
  \end{center}%
}

\makeatletter
\newcommand{\institute}[1]{\newcommand{\@institute}{#1}}

\renewcommand\textsl{\textcolor{blue}}

\begin{document}
\title{Quantifying the Causal Operational Determinants of Service Reliability in Urban Rail Transit: Evidence from Panel Double/Debiased Machine Learning}
\author[]{Ying Yao}
\author[]{Nan Zhang}
\author[]{Daniel J. Graham$^*$}
\affil[]{Department of Civil Engineering, Imperial College London, London, SW7 2AZ, UK.\\ $^*$ Corresponding Author: \texttt{d.j.graham@imperial.ac.uk}} 
%

\date{}

\maketitle

\begin{abstract}
Urban rail transit reliability is a critical measure of system performance, yet its causal determinants remain poorly quantified due to high-dimensional and interdependent influencing factors. This study investigates reliability patterns across 46 international metro operators between 1994 and 2024 using the CoMET benchmarking database, incorporating more than 90 candidate variables spanning technical, operational, financial, environmental, and macroeconomic conditions. Based on domain knowledge, literature synthesis, and variable construction, four operational determinants are designed to capture three mechanisms: demand pressure, service supply, and demand–supply imbalance, while the remaining variables are screened and incorporated as confounders where theoretically appropriate.

Double/Debiased Machine Learning (DML) adapted for panel data is introduced to urban rail reliability analysis to quantify the net causal effects of these determinants under complex and nonlinear relationships. The framework combines flexible machine learning with panel fixed or random effects within-operator temporal variation, reducing bias from high-dimensional confounding, model misspecification, and unobserved operator heterogeneity.

The results identify three distinct operational mechanisms. Higher passenger demand intensity increases incident rates by 0.38\% ($p<0.001$). On the supply side, greater fleet supply adequacy and car-based operational intensity reduce incident rates by 0.52\% ($p<0.05$) and 0.80\% ($p<0.01$), respectively. Capacity utilization, which reflects the imbalance between demand and available supply, increases incident rates by 0.49\% ($p<0.001$). These findings show that metro reliability depends not only on the level of demand or supply alone, but also on whether service provision keeps pace with passenger demand.
\end{abstract}
.

\section{Introduction}

Urban rail transit (URT) reliability concerns a system's ability to maintain service during disruptions while preserving network connectivity and accommodating passenger flows without substantial increases in travel impedance \citep{Liu2020}. It therefore extends beyond the physical availability of stations and lines and should be understood as a multidimensional concept involving infrastructure, operations, passenger demand, and network-wide interactions \citep{Shen2026}. Because rail lines are operationally interconnected, incidents at individual stations or sections may propagate through interchange stations and develop into system-wide disruptions, with wider consequences for accessibility, productivity, and urban resilience \citep{Cats2020}. Within this broader reliability literature, the present study focuses specifically on operational reliability at the annual operator level, measured by the frequency of incidents causing delays of five minutes or longer, rather than on structural connectivity or post-disruption recovery.

These challenges are becoming increasingly important as metro systems accommodate growing passenger demand within constrained infrastructure. A central operational question is how intensively available capacity can be used before incident risk rises, and how much service buffer is required to maintain acceptable reliability. Addressing this question requires credible evidence on how incident rates respond to passenger demand, service supply, and the balance between them.

Highly disaggregated automatic vehicle location and smart-card data can link short-term crowding and loading conditions to disruptions within individual metro systems. However, such data are rarely available consistently across systems and over long periods, limiting the generalisability of findings. Individual systems may also operate within relatively narrow ranges of demand and supply, making it difficult to identify effects across the broader operational spectrum. More importantly, operational conditions are not randomly assigned. Passenger demand and service provision reflect network design, planning decisions, operating practices, and historical investment, many of which may also influence reliability. Simple associations may therefore confound the effects of operational determinants with differences in system characteristics and operating environments.

This study uses a unique longitudinal dataset covering 46 international URT operators between 1994 and 2024 to estimate the causal effects of four operational determinants on URT reliability. 81 candidate variables describe reliability outcomes, passenger demand, service supply, infrastructure, operational resources, safety, and financial performance. Drawing on domain knowledge and the literature, four operational determinants are constructed to represent three mechanisms: demand pressure, service supply, and demand--supply imbalance. The remaining variables are evaluated as potential treatment-specific confounders and included where theoretically justified.

The panel structure supports a causal strategy based on within-operator variation over time. Fixed-effects adjustment controls for additive time-invariant heterogeneity across metro systems, including persistent differences in governance, infrastructure, network configuration, and operating culture. Treatment-specific covariate adjustment addresses observed time-varying factors that may jointly influence operational conditions and reliability. Given this causal model and the corresponding adjustment sets, Double/Debiased Machine Learning (DML) is used to estimate the target effects with flexible adjustment for high-dimensional and potentially nonlinear confounding \citep{Chernozhukov2018}. Through Neyman orthogonalization and cross-fitting, DML reduces sensitivity to regularization, overfitting, and misspecification of the nuisance relationships between confounders, treatment, and outcome. Causal identification nevertheless depends on the validity of the underlying adjustment sets and maintained identification assumptions rather than on DML itself.

This paper makes three main contributions. First, conceptually, it develops a theory-informed framework distinguishing demand pressure, service supply, and demand--supply imbalance as interrelated mechanisms affecting URT reliability. Second, empirically, it provides, to our knowledge, the first causal analysis of metro reliability using an international longitudinal panel constructed from the CoMET benchmarking database. Third, methodologically, it adapts DML to an international urban rail benchmarking context by combining flexible nuisance-function estimation with fixed-effects transformation to address high-dimensional observed confounding and time-invariant operator heterogeneity.

The remainder of the paper is organized as follows. Section~2 reviews the literature. Section~3 describes the data and variable construction. Section~4 presents the methods. Section~5 reports the results, and Section~6 concludes.

\section{Literature review}

\subsection{Concepts of URT reliability}
Urban rail transit (URT) reliability has been conceptualized from multiple perspectives, reflecting the diverse ways in which service performance can deteriorate under disruptions. Existing studies generally distinguish three complementary dimensions: structural reliability, functional reliability, and overall network robustness or resilience \citep{Cats2020,Li2019Resilience,Wang2024Urban}. Structural reliability concerns the physical connectivity of the network and the availability of feasible paths between origin–destination pairs \citep{Bell1997,Wakabayashi1992,Zhang2015}. Functional reliability focuses on service quality experienced by passengers, commonly measured through travel time variability, accessibility, or journey reliability \citep{Asakura1991,Guo2012,Yang2014}. Capacity reliability evaluates whether the network can continue accommodating passenger demand under degraded operating conditions by examining residual capacity or congestion resilience \citep{Chen1999,Shariat2013,Chen2013,Liu2020}. Together, these perspectives demonstrate that URT reliability extends beyond infrastructure availability to encompass passenger experience, operational performance, and system resilience through interactions between network structure, operations, and disruption management under both normal and disrupted operating conditions \citep{Zhang2022Quantitative,Pan2024,Lin2024Metro,Wang2024Urban}.

\subsection{Network-based robustness}
Network-based robustness analysis has been widely applied to evaluate the vulnerability of transportation systems by sequentially removing stations or track segments, often guided by metrics such as degree centrality and betweenness centrality. Results consistently indicate that attacks based on betweenness centrality cause the most severe disruptions \citep{Cats2020}. However, while these network-centric approaches offer valuable insights into structural vulnerabilities, they typically overlook demand dynamics, operational management, maintenance conditions, and performance degradation caused by congestion \citep{Zhang2022Quantitative,Wang2024Urban,Consilvio2024A}.

\subsection{Operational determinants of reliability}
Recent studies have shown that URT reliability is not determined by network topology alone, but emerges from the interaction between network structure, passenger demand, operating strategies, and disturbance management throughout the disruption process \citep{Ma2022,Pan2024}. Reliability is therefore inherently dynamic rather than static, varying across operating conditions and time periods. In particular, reliability generally deteriorates during peak demand periods and under disruptions affecting transfer functions, network capacity, or recovery processes \citep{Zhu2025Measuring,Liu2020Connectivity,Zhu2025Percolation,Chen2021Vulnerability}.

Passenger demand has consistently been identified as one of the most important operational determinants of URT reliability \citep{Besinovic2021A,Zhang2024Dynamic}. Peak-hour demand increases operational pressure through higher passenger volumes, longer dwell times, reduced recovery margins, and transfer-station crowding, resulting in substantially poorer reliability than under off-peak conditions \citep{Liu2020,Pan2024,Zhu2025Measuring,Chen2021Vulnerability,Zhang2024Dynamic}. Accordingly, several recent reliability assessment frameworks explicitly incorporate temporal demand variation and passenger flow control into reliability evaluation \citep{Besinovic2021A}. However, passenger demand alone provides only a partial description of operational pressure, as reliability is also influenced by how effectively available capacity accommodates demand across trains, stations, and network sections.

Capacity utilization has consequently emerged as an increasingly important operational indicator linking passenger demand with service supply. Rather than representing passenger volume alone, capacity utilization reflects the extent to which available capacity at the train, station, and network levels is consumed by passenger demand \citep{Peftitsi2021Evaluating,Zhang2024Dynamic,Liang2024Data-driven}. Existing studies show that utilization is inherently dynamic, varying across time, locations, and vehicle cars, with local bottlenecks often occurring even when system-wide capacity remains available \citep{Peftitsi2021Evaluating,Li2024Scheduling,Zhao2018Location}. Recent research therefore emphasizes that demand–capacity mismatch, rather than demand alone, is a primary contributor to crowding, operational stress, and service degradation \citep{Liang2024Data-driven,Yang2025Assessment}.

Service supply and infrastructure characteristics complement capacity utilization  by determining the amount of operational capacity available to accommodate passenger demand. Service supply and infrastructure characteristics, including transfer facilities and asset condition, represent another major group of operational determinants. Capacity reductions caused by disrupted links or degraded infrastructure significantly reduce network performance, particularly under growing passenger demand \citep{Liu2020Effects,Zhang2021Metro}. Critical transfer stations and highly connected links have also been shown to disproportionately influence system reliability because disturbances at these locations propagate rapidly across the network \citep{Chen2022Resilience,Pan2024,Zhang2024Dynamic}. Beyond network operations, maintenance capability and asset management have likewise been recognized as important contributors to long-term operational reliability through maintenance prioritization based on both asset condition and passenger impact \citep{Sun2020Maintenance,Consilvio2024A}.

International benchmarking has become an established approach to compare metro performance across operators and facilitate organizational learning, while emphasizing the need to adjust for contextual differences rather than relying on raw performance comparisons \citep{Graham2022Model-based,Zope2019Benchmarking}.

More recently, several studies have begun to examine operational reliability using observed system performance rather than network topology. \citet{Melo2011} analyzed incident rates across 15 international metro systems and more than 40 metro lines, using incidents per car-kilometer as the reliability measure. Passenger journeys were included as one explanatory variable, demonstrating that operational characteristics can partially explain differences in reliability across metro systems. However, international benchmarking datasets have so far been used primarily for descriptive performance evaluation, with comparatively limited attention devoted to causal inference on operational reliability.

Nevertheless, operational determinants have largely been investigated separately, despite their inherent interdependence in real-world metro operations. A unified empirical framework that jointly evaluates passenger demand, service supply, and capacity utilization while accounting for high-dimensional confounding is still lacking.

\subsection{Causal inference in transportation}
More broadly, causal inference has received increasing attention in transportation research as observational data have become more widely available. Rather than relying on statistical associations alone, recent studies have adopted quasi-experimental and counterfactual frameworks to estimate the causal effects of transport policies and operational interventions, including difference-in-differences, instrumental variables, synthetic control, and matching methods \citep{Kim2016Quasi-Experimental,Cham2024Quasi-Experimental,Xu2017Generalized}. More recently, machine-learning-based causal estimators have been introduced to accommodate complex nonlinear relationships and heterogeneous treatment effects in transportation systems \citep{Zhou2024Machine,Zhang2025Ex-post,Graham2025Causal}. However, these developments have primarily focused on travel behavior, policy evaluation, and environmental impacts, with comparatively limited application to urban rail operational reliability.

\subsection{Research gap}

Despite these advances in both URT reliability analysis and transportation causal inference, several important gaps remain. First, existing studies predominantly examine either network topology or isolated operational indicators, rather than providing a unified framework that jointly evaluates demand pressure, service supply, and capacity utilization. Second, most empirical analyses rely on conventional regression methods, which are vulnerable to multicollinearity, nonlinear relationships, and high-dimensional confounding inherent in URT operational data. Third, causal evidence remains limited because operational variables such as passenger demand and service provision are jointly determined by long-term planning decisions and network characteristics, making simple associations difficult to interpret causally. To the best of our knowledge, no previous URT reliability study has formally addressed these causal identification issues within the potential outcomes framework. These limitations motivate the development of a causal framework capable of quantifying the independent effects of multiple operational determinants while accounting for complex confounding structures. The proposed framework addresses these gaps by integrating international benchmarking data with panel causal inference and machine learning.

\section{Data}
\subsection{CoMET benchmarking database}
The CoMET programme is an international metro benchmarking initiative in which participating operators voluntarily report harmonized operational and performance indicators on an annual basis. The analysis uses the CoMET international benchmarking database, comprising an unbalanced panel of 46 metro operators between 1994 and 2024. The database contains more than 900 operational indicators covering passenger demand, service supply, infrastructure, fleet characteristics, maintenance, labor inputs, financial performance, safety outcomes, and macroeconomic conditions. Rather than analyzing all variables directly, the database is organized into a causal framework based on domain knowledge and the operational reliability literature.

\subsection{Variable construction}
\subsubsection{Outcome variable}

Operational reliability is measured by the number of operational incidents causing delays of five minutes or longer per million actual revenue train-kilometers. Unlike the incident measure adopted by \citet{Melo2011}, which is normalized by car-kilometers, the present study uses train-kilometers to reflect the volume of train service operated. The outcome variable is defined as

\[
\text{Incident Rate}
=
\frac{\text{Total Number of Incidents Causing Delays of 5+ Minutes}}
{\text{Million Actual Revenue Train-km}}.
\]

Normalizing incidents by train-kilometers accounts for differences in service volume across metro operators and over time, thereby improving comparability between systems of different sizes. This indicator is drawn from the CoMET benchmarking framework and represents the frequency of reliability-related incidents relative to the amount of train service delivered.

\subsubsection{Operational determinants}

Following the conceptual framework established in Section~2, four operational determinants are constructed to represent three complementary operational mechanisms: passenger demand, service supply, and demand–supply imbalance. Two service-supply indicators are included because service provision can be characterized relative to either the operated route length or the operated network length. Table~\ref{tab:treatments} summarizes the definition, formulation, and conceptual interpretation of each operational determinant.

\begin{table}[htbp]
\centering
\caption{Operational determinants used in the causal analysis.}
\label{tab:treatments}
\small
\renewcommand{\arraystretch}{1.25}

\begin{tabular}{p{4.3cm}p{5.3cm}p{5.9cm}}
\toprule
\textbf{Determinant} &
\textbf{Definition} &
\textbf{CoMET formulation} \\
\midrule

Annual Network Passenger Intensity (ANPI) &
\textbf{Demand.} System-level passenger demand normalized by the operated network length. &
Passenger km / Network Length -- Operated \\

Fleet Supply Adequacy (FSA) &
\textbf{Supply.} Revenue service supply normalized by the operated route length. &
Car km -- Actual Revenue Operating / Route Length -- Operated \\

Car-based Operational Intensity (COI) &
\textbf{Supply.} Revenue service supply normalized by the operated network length. &
Car km -- Actual Revenue Operating / Network Length -- Operated \\

Capacity Utilization (CU) &
\textbf{Demand--supply imbalance.} Passenger demand relative to available operating capacity. &
Passenger km / Standardized Capacity km \\

\bottomrule
\end{tabular}
\vspace{0.3em}

\begin{minipage}{0.96\textwidth}
\footnotesize
\textit{Note:} Throughout the remainder of the paper, the four operational determinants are abbreviated as ANPI, FSA, COI, and CU.
\end{minipage}
\end{table}

For each determinant, the hypothetical intervention is defined as a 1\% proportional increase in the corresponding operational index. This represents a 1\% increase in annual passenger-km per operated network-km for ANPI, actual revenue car-km per operated route-km for FSA, actual revenue car-km per operated network-km for COI, and passenger-km relative to standardized capacity-km for CU. Because each treatment is a ratio, the change may arise through its numerator, denominator, or both. The estimand therefore pertains to the composite index and does not identify the separate causal effects of its components.

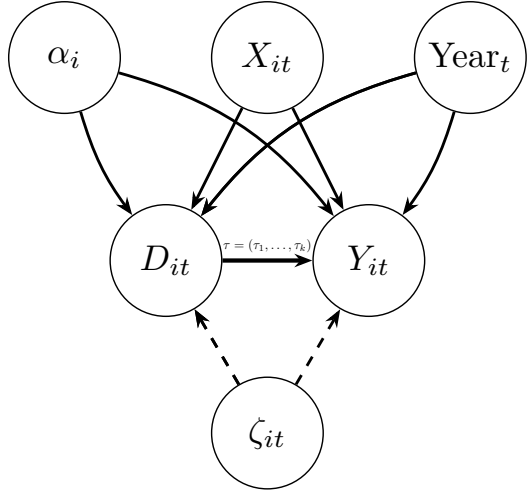
\begin{figure}[t]
\centering
\resizebox{0.45\linewidth}{!}{%
\begin{tikzpicture}[
  every node/.style={
    circle,
    draw,
    minimum size=1.1cm,
    inner sep=2pt,
    font=\small
  },
  arr/.style={-{Stealth[length=5pt]}, thick},
  causal/.style={-{Stealth[length=5pt]}, very thick},
  latent/.style={-{Stealth[length=5pt]}, thick, dashed}
]
\node (A) at (0,2) {$\alpha_i$};
\node (X) at (2,2) {$X_{it}$};
\node (T) at (4,2) {$\mathrm{Year}_t$};
\node (D) at (1,0) {$D_{it}$};
\node (Y) at (3,0) {$Y_{it}$};
\node (U) at (2,-1.7) {$\zeta_{it}$};
\draw[arr] (A) to[bend right=10] (D);
\draw[arr] (A) to[bend left=18] (Y);
\draw[arr] (X) -- (D);
\draw[arr] (X) -- (Y);
\draw[arr] (T) to[bend left=10] (Y);
\draw[arr] (T) to[bend right=18] (D);
\draw[arr] (T) to[bend right=18] (D);
\draw[causal] (D) -- node[draw=none, fill=none, midway, above=-12pt]{\scalebox{0.37}{$\tau=(\tau_1,\ldots,\tau_k)$}} (Y);
\draw[latent] (U) -- (D);
\draw[latent] (U) -- (D);
\draw[latent] (U) -- (Y);
\end{tikzpicture}
}
\caption{Causal structure underlying the Panel DML analysis. 
}
\label{fig:dag}
\end{figure}

Figure~\ref{fig:dag} summarizes the causal structure underlying the empirical analysis. For each operational determinant $D_{it}$, the adjustment set $X_{it}$ contains treatment-specific pre-treatment covariates that may jointly affect the determinant and the reliability outcome $Y_{it}$ (e.g. confounders). Time-invariant operator characteristics are represented by $\alpha_i$ and controlled through the fixed-effects transformation, while $\mathrm{Year}_t$ accounts for common temporal variation, and $\zeta_{it}$ allows for the possibility of unobserved confounding. The parameter of interest corresponds to the causal path from $D_{it}$ to $Y_{it}$.
Identification requires that, conditional on $X_{it}$, $\alpha_i$, and $\mathrm{Year}_t$, no remaining time-varying unobserved factor jointly determines the operational determinant and the reliability outcome. Accordingly, the adjustment sets include pre-treatment confounders but exclude variables that may be consequences of the treatment.

\subsubsection{Confounder selection}
Treatment-specific adjustment sets are selected based on the causal structure of each determinant. Table~\ref{tab:confounders} reports the parsimonious sets used in the adjusted panel regressions, while Table~\ref{tab:ddml_covariates} reports the broader covariate sets used for DML nuisance-function estimation.
\begin{table}[htbp]
\centering
\caption{Treatment-specific confounders used in the adjusted panel regression models.}
\label{tab:confounders}
\small
\renewcommand{\arraystretch}{1.15}
\begin{tabular}{p{6.4cm}cccc}
\toprule
\textbf{Confounder} &
\textbf{ANPI} &
\textbf{FSA} &
\textbf{COI} &
\textbf{CU} \\
\midrule
\multicolumn{5}{l}{\textbf{Operational scale and demand}}\\
Capacity Utilization &
&
\cmark &
\cmark &
\\
Standardized Capacity km &
\cmark &
&
&
\\
Network Length -- Operated &
&
&
&
\cmark \\
\addlinespace
\multicolumn{5}{l}{\textbf{Fleet characteristics}}\\
Cars per Train &
\cmark &
\cmark &
\cmark &
\cmark \\
Fleet -- Cars which are Available &
\cmark &
&
\cmark &
\cmark \\
Average Car Length &
&
&
\cmark &
\cmark \\
Car Seating Capacity &
\cmark &
\cmark &
\cmark &
\cmark \\
Design Space for Standing Passengers &
\cmark &
&
&
\cmark \\
\addlinespace
\multicolumn{5}{l}{\textbf{Operational characteristics}}\\
Average Commercial Speed &
\cmark &
&
\cmark &
\cmark \\
Total Own Labor Costs &
\cmark &
\cmark &
\cmark &
\cmark \\
\bottomrule
\end{tabular}
\vspace{0.3em}

\begin{minipage}{0.7\textwidth}
\footnotesize
\textit{Note:}
ANPI = Annual Network Passenger Intensity;
FSA = Fleet Supply Adequacy;
COI = Car-based Operational Intensity;
CU = Capacity Utilization.
\end{minipage}
\end{table}

\subsection{Data preprocessing}
The original CoMET database contains annual observations for 46 metro operators together with more than 900 raw operational indicators collected from the KPI and Profile datasets. These records were first organized into an unbalanced panel covering the period 1994--2024. Operational reliability was measured using the incident rate defined in Section~3.2.1, while the remaining variables served as candidate determinants or potential confounders.

To improve data quality, indicators with more than 70\% missing observations were first excluded, reducing the initial pool of over 900 raw indicators to 81 candidate variables. The remaining dataset was then restricted to observations with complete information on the outcome and the corresponding treatment-specific adjustment set. This sequential procedure preserves variables with sufficient temporal coverage while avoiding excessive loss of observations caused by high-dimensional missingness. Because reporting completeness differs across operators and years, the final analytical dataset remains an unbalanced panel, with the sample size varying slightly across model specifications according to the availability of the corresponding treatment and confounder variables.

For the linear and panel benchmark models, estimation is based on the complete-case sample obtained through the filtering procedure described above. In contrast, the DML framework estimates nuisance functions using XGBoost, which accommodates missing covariate values without requiring imputation. Consequently, observations are retained provided that the outcome and treatment variables are observed and each metro operator contributes at least three observations, while any remaining missingness in the confounders is handled directly by the ML algorithm. This increases the effective sample size from approximately 300 complete cases to around 500 observations per treatment, allowing the DML models to exploit substantially more of the available longitudinal information while avoiding unnecessary sample loss due to missing confounders. For the Panel DML specifications, the outcome, treatment, and covariates are transformed using operator-specific within-group demeaning for FE or quasi-demeaning for RE, with operator means calculated from the available observations only and any remaining missing covariate values passed directly to XGBoost.

\section{Methods}
\subsection{Naive OLS}
As a baseline, naive OLS models were estimated separately for each operational determinant after applying logarithmic transformations to both the outcome and treatment variables, allowing the estimated coefficients to be interpreted as elasticities while reducing distributional skewness. For each determinant, the following polynomial specification was estimated:

 \begin{equation}
\log Y_{it}
=
\alpha
+\theta_1\log D_{it}
+\theta_2(\log D_{it})^2
+\theta_3(\log D_{it})^3
+\delta\,\mathrm{Year}_t
+\varepsilon_{it},
\label{eq:ols_poly}
\end{equation}

where lower-order models were obtained by omitting the higher-order polynomial terms. Linear, quadratic, and cubic specifications were compared, and the preferred functional form for each determinant was selected using the Akaike Information Criterion (AIC).

\subsection{Panel methods}
Building on the baseline OLS analysis, the panel models retain the logarithmic transformation of both the outcome and determinants. The treatment-specific functional form selected by the AIC is estimated using both fixed-effects (FE) and random-effects (RE) panel models:

\begin{equation}
\log Y_{it}
=
\alpha_i+\theta_1\log D_{it}
+\theta_2(\log D_{it})^2
+\theta_3(\log D_{it})^3
+\delta\,\mathrm{Year}_t
+\varepsilon_{it},
\end{equation}

where $\alpha_i$ denotes the operator-specific effect, treated as fixed under FE and as random under RE, with $\alpha_i \sim \mathcal{N}(0,\sigma_{\alpha}^{2})$. Model selection is guided by the Hausman test \citep{Hausman1978}. Under the null hypothesis, $\alpha_i$ is uncorrelated with the regressors, so RE is consistent and efficient; rejection implies that RE is inconsistent and FE is preferred. Conceptually, FE also controls for additive time-invariant operator heterogeneity correlated with both treatment and outcome, whereas RE assumes such heterogeneity is independent of $D_{it}$. FE is therefore preferred when residual time-invariant confounding is plausible.

To compare polynomial specifications, effects are reported as average marginal effects (AMEs), interpreted as average elasticities:

\begin{equation}
\text{AME}
=
\frac{1}{N}
\sum_{i=1}^{n}
\sum_{t=1}^{T_i}
\left[
\theta_1
+2\theta_2\log D_{it}
+3\theta_3(\log D_{it})^2
\right],
\end{equation}

where $N=\sum_{i=1}^{n}T_i$. Under the linear specification, the AME reduces to $\theta_1$.

Nonlinearity is assessed using Wald tests \citep{Wald1943,Engle1984}. For the quadratic model, $H_0:\theta_2=0$; for the cubic model, $H_0:\theta_2=\theta_3=0$. A small $p$-value indicates evidence of nonlinearity beyond the linear specification.

\subsection{Double / Debiased Machine Learning (DML)}

\subsubsection{DML estimation under high-dimensional confounding}

Double (or Debiased) Machine Learning (DML) \citep[e.g.][]{Chernozhukov2018} is used to estimate low-dimensional causal parameters in the presence of high-dimensional or complex nuisance functions. Given the treatment-specific causal model and adjustment sets established in the Data section, DML allows the target causal effects to be estimated while flexibly adjusting for high-dimensional and potentially nonlinear observed confounding.

The role of DML is estimation rather than causal identification. Identification depends on the conditional independence and overlap assumptions supported by the treatment-specific selection of confounders. Conditional on these assumptions, DML reduces sensitivity to misspecification of the nuisance relationships between the confounders, treatment, and outcome. It does not compensate for omitted confounders, inappropriate adjustment variables, or other violations of the underlying causal model.

DML estimates the nuisance functions using flexible machine-learning algorithms. Directly substituting these estimates into the estimating equation may nevertheless introduce regularization and overfitting bias. DML addresses these problems through Neyman orthogonalization and cross-fitting. The orthogonal score makes estimation of the target parameter locally insensitive to small errors in the nuisance functions, while cross-fitting estimates the nuisance functions and evaluates the score on separate subsets of the data.

\subsubsection{DML for causal estimands}

DML is used to estimate the target causal effect while flexibly adjusting for high-dimensional observed confounding. Causal identification relies on the conditional independence assumption (CIA) and sufficient overlap in the conditional treatment distribution:

\begin{equation}
Y_{it}(d) \perp\!\!\!\perp D_{it} \mid X^{(D)}_{it}, \qquad 0 < f_{D\mid X^{(D)}}(d \mid x) < \infty,
\end{equation}

for all relevant values of $d$ and $x$, where $X_{it}^{(D)}$ denotes the treatment-specific adjustment set. These assumptions are maintained on the basis of the causal model and treatment-specific confounder-selection procedure rather than established by the DML estimator itself.

For notational simplicity, $Y_{it}$ and $D_{it}$ denote the logarithmically transformed outcome and operational determinant used in estimation. The analysis adopts the partially linear regression (PLR) model

\begin{equation}
Y_{it} = D_{it}\theta_0 + g_0\!\left(X^{(D)}_{it}\right) + \varepsilon_{it}, \qquad \mathbb{E}\!\left[\varepsilon_{it} \mid D_{it}, X^{(D)}_{it}\right] = 0,
\end{equation}

where $g_0(\cdot)$ is an unknown nuisance function and $\theta_0$ is the target causal parameter. For continuous treatments, $\theta_0$ denotes the average partial causal effect and, under the log--log specification, the causal elasticity of the incident rate with respect to the operational determinant.

Define the nuisance functions
\begin{equation}
\ell_0(X)
=
\mathbb{E}[Y\mid X],
\qquad
m_0(X)
=
\mathbb{E}[D\mid X],
\end{equation}
representing the conditional outcome and treatment regressions, respectively. The Neyman-orthogonal score for the PLR model is
\begin{equation}
\psi(W;\theta,\eta)
=
\bigl(D-m(X)\bigr)
\left[
Y-\ell(X)
-
\bigl(D-m(X)\bigr)\theta
\right],
\end{equation}
where $W=(Y,D,X)$ and $\eta=(\ell,m)$. The DML estimator is obtained by solving the sample analogue of the orthogonal moment condition using cross-fitted estimates of the nuisance functions.

In this study, $\ell(X)$ and $m(X)$ are estimated using XGBoost, allowing flexible adjustment for high-dimensional and potentially nonlinear confounding relationships. Orthogonalization and cross-fitting reduce the sensitivity of the estimated treatment effect to regularization, overfitting, and misspecification of these nuisance functions. The causal interpretation nevertheless remains conditional on the validity of the treatment-specific adjustment sets and the identification assumptions stated above.

\subsubsection{Panel DML}

The baseline DML specification treats observations as independent and therefore does not explicitly account for persistent differences across metro operators. Such differences may reflect network configuration, rolling-stock characteristics, operating practices, governance structures, or other institutional conditions that are correlated with both the operational determinants and reliability. To account for this time-invariant operator heterogeneity, we apply the within-group Panel DML approach, in which the variables are demeaned at the operator level before flexible estimation of the nuisance functions \citep{Clarke2026Double}.

Consider the panel PLR model
\begin{equation}
Y_{it}
=
D_{it}\theta_0
+
g_0\!\left(X_{it}^{(D)}\right)
+
\alpha_i
+
\varepsilon_{it},
\end{equation}
where $\alpha_i$ captures time-invariant operator-specific heterogeneity and $X_{it}^{(D)}$ contains the treatment-specific observed confounders and Year. Based on the Hausman tests reported above, the fixed-effects specification is adopted for all four operational determinants. The outcome, treatment, and nuisance covariates are therefore transformed using operator-specific within-group demeaning:
\begin{equation}
\widetilde{Z}_{it}
=
Z_{it}
-
\overline{Z}_{i},
\qquad
Z\in\{Y,D,X\},
\end{equation}
where $\overline{Z}_{i}$ denotes the operator-specific mean calculated from the available observations.

The within transformation complements, rather than replaces, adjustment for the observed treatment-specific confounders. The former removes time-invariant operator heterogeneity, whereas the latter supports conditional identification with respect to observed time-varying confounding. DML is subsequently applied to the within-transformed outcome, treatment, and covariates.

Let the cross-fitted residuals of the transformed outcome and treatment be defined as
\begin{align}
\widehat{U}_{it}
&=
\widetilde{Y}_{it}
-
\widehat{\ell}^{\,(-k(i,t))}
\left(
\widetilde{X}_{it}^{(D)}
\right),
\\
\widehat{V}_{it}
&=
\widetilde{D}_{it}
-
\widehat{m}^{\,(-k(i,t))}
\left(
\widetilde{X}_{it}^{(D)}
\right),
\end{align}
where $k(i,t)$ denotes the fold containing observation $(i,t)$, and
$\widehat{\ell}^{\,(-k)}$ and $\widehat{m}^{\,(-k)}$ are estimated using observations outside that fold. The Panel DML estimator is then given by
\begin{equation}
\widehat{\theta}
=
\frac{
\displaystyle
\sum_{i,t}
\widehat{V}_{it}\widehat{U}_{it}
}{
\displaystyle
\sum_{i,t}
\widehat{V}_{it}^{\,2}
}.
\end{equation}

Both nuisance functions are estimated using an ensemble of two XGBoost learners with different hyperparameter settings. Ensemble weights are selected through five-fold cross-validation, and five-fold cross-fitting is conducted at the observation level. Because the outcome and treatment variables are expressed in logarithms and transformed within operators, the resulting coefficients are interpreted as within-operator causal elasticities, conditional on the treatment-specific adjustment sets, the maintained identification assumptions, and the within-group approximation.

\section{Results \& Discussion}

\subsection{Baseline OLS analysis}

Among the four operational determinants, ANPI was best represented by a linear specification, CU and COI by quadratic specifications, and FSA by a cubic specification. However, adjusted $R^2$ values were generally below 0.30, indicating that the baseline OLS models explained only a limited share of the variation in operational reliability. These estimates therefore serve primarily as descriptive benchmarks.

\subsection{Panel regression}

Table~\ref{tab:panel_results} reports the FE and RE estimates and the corresponding Hausman tests. The tests favor fixed effects for all four determinants.

\begin{table}[htbp]
\centering
\caption{Panel regression results for the four operational determinants.}
\label{tab:panel_results}
\small
\renewcommand{\arraystretch}{1.15}
\setlength{\tabcolsep}{4pt}

\begin{tabular}{llccc ccc cc}
\toprule

&
&
\multicolumn{3}{c}{\textbf{FE}}
&
\multicolumn{3}{c}{\textbf{RE}}
&
\multicolumn{2}{c}{\textbf{Hausman}}\\

\cmidrule(lr){3-5}
\cmidrule(lr){6-8}
\cmidrule(lr){9-10}

\textbf{Determinant}
&
\textbf{Form}
&
\textbf{AME}
&
\textbf{Adj.\ $R^2$}
&
\textbf{$p$}
&
\textbf{AME}
&
\textbf{Adj.\ $R^2$}
&
\textbf{$p$}
&
$\chi^2$
&
\textbf{Model}
\\

\midrule

ANPI & Lin.  & +0.346 & -0.014 & $<0.001$ & +0.265 & 0.045 & 0.003 & 16.92*** & FE \\

FSA  & Cub.  & -0.330 & 0.000 & 0.852 & -0.298 & 0.074 & 0.694 & 12.81* & FE \\

COI  & Quad. & -0.327 & -0.046 & 0.328 & -0.432 & 0.030 & 0.165 & 38.76*** & FE \\

CU   & Quad. & +0.875 & 0.060 & $<0.001$ & +0.801 & 0.108 & $<0.001$ & 28.61*** & FE \\

\bottomrule
\end{tabular}
\end{table}

Relative to OLS, the FE estimates change substantially. In particular, the estimated effects of ANPI and CU become positive, suggesting that cross-sectional differences between metro systems obscure their within-operator relationships with reliability. The effects of FSA and COI are not statistically distinguishable from zero at this stage. The generally low within $R^2$ values indicate that controlling only for time-invariant operator heterogeneity remains insufficient, motivating further adjustment for observed confounders.

\subsection{Panel regression with observed confounders}

\begin{figure}[!t]
    \centering
    \includegraphics[width=0.75\linewidth]{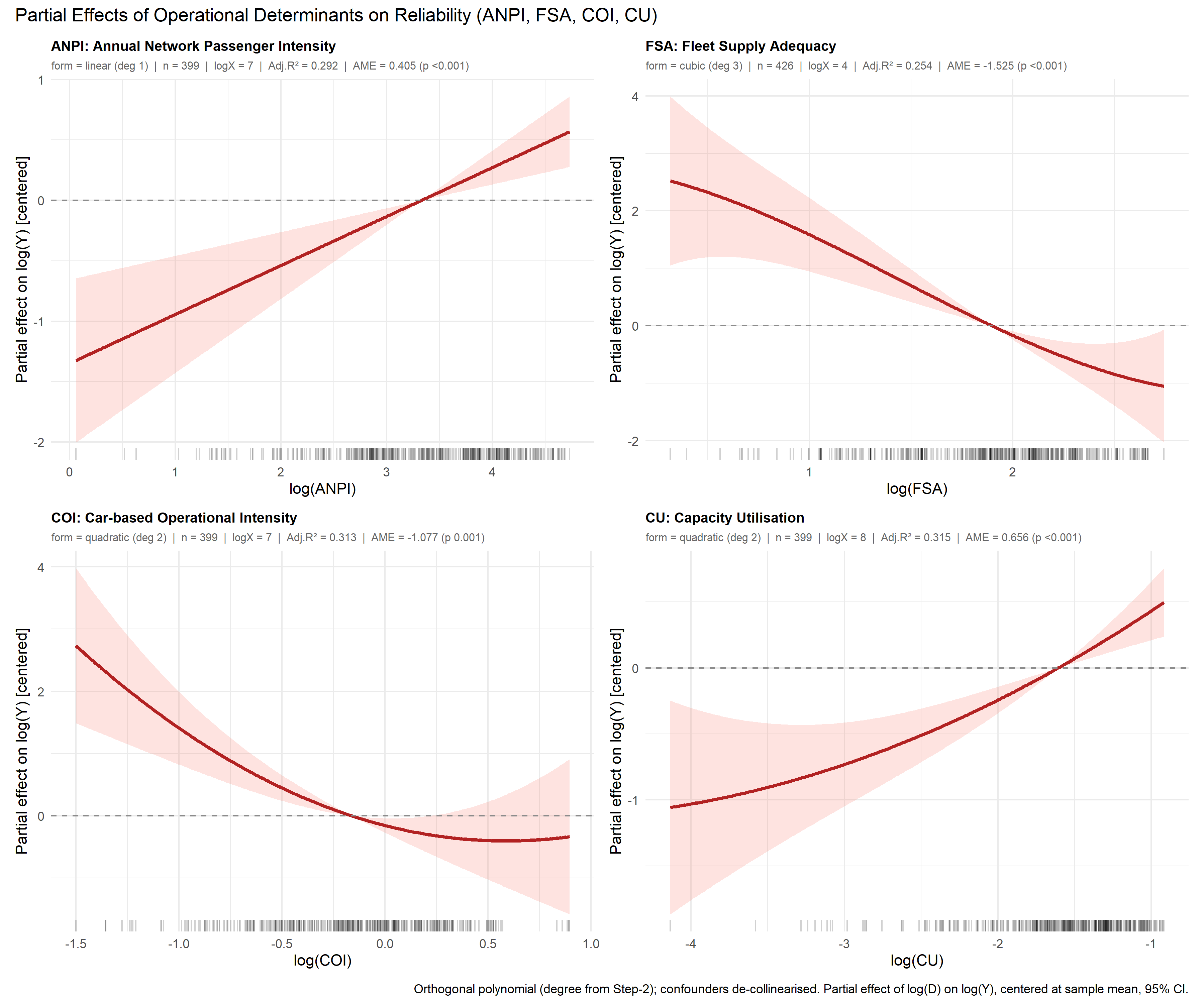}
    \caption{Partial effects of operational determinants on reliability.}
    \label{fig:panel_confounders}
\end{figure}

The preferred panel specifications were re-estimated with the treatment-specific confounders described in the Data section. Table~\ref{tab:panel_confounder_results} reports the resulting average marginal effects (AMEs), and Fig.~\ref{fig:panel_confounders} shows the corresponding partial-effect curves.

\begin{table}[htbp]
\centering
\caption{Panel regression results after adjustment for observed confounders.}
\label{tab:panel_confounder_results}
\small
\renewcommand{\arraystretch}{1.15}

\begin{tabular}{lccccccccc}
\toprule
\textbf{Determinant}
&
\textbf{$N$}
&
\textbf{Conf.}
&
\textbf{AME}
&
\textbf{SE}
&
\textbf{$p$}
&
\textbf{Adj.\ $R^2$}
&
\textbf{Higher-order}
&
\textbf{$p^*$}
\\
\midrule

ANPI
&
399
&
7
&
+0.405
&
0.106
&
$<0.001$
&
0.292
&
Linear
&
--

\\

FSA
&
426
&
4
&
$-1.525$
&
0.331
&
$<0.001$
&
0.254
&
Cubic
&
0.721

\\

COI
&
399
&
7
&
$-1.077$
&
0.334
&
0.001
&
0.313
&
Quadratic
&
0.050

\\

CU
&
399
&
8
&
+0.656
&
0.149
&
$<0.001$
&
0.315
&
Quadratic
&
0.265

\\

\bottomrule
\end{tabular}

\vspace{0.35em}

\footnotesize
\textit{Notes.}
$N$ is the sample size; Conf.\ denotes the number of retained confounders; AME denotes the average marginal effect. For FSA, $p^*$ is based on a Wald joint test of the quadratic and cubic terms.

\end{table}

Confounder adjustment increases the adjusted $R^2$ to 0.254--0.315. The estimated AMEs are +0.405 for ANPI, $-1.525$ for FSA, $-1.077$ for COI, and +0.656 for CU. The supply-related effects (FSA and COI) become considerably stronger, whereas the elasticity of CU decreases slightly; the demand and capacity-utilization effects nonetheless remain positive.

The higher-order terms receive little support after adjustment. The quadratic term is insignificant for CU ($p*=0.265$) and only marginally significant for COI ($p*=0.050$), while the quadratic and cubic terms for FSA are jointly insignificant ($p*=0.721$). The apparent nonlinearities in the baseline models therefore weaken once observed confounding is addressed. Nevertheless, the panel regressions remain dependent on parametric adjustment, motivating the more flexible DML specification.

\subsection{DML estimation}

\subsubsection{Baseline DML}

DML retains a broader set of theoretically relevant adjustment variables and estimates the nuisance functions using XGBoost. The treatment-specific covariate sets are reported in Table~\ref{tab:ddml_covariates}.

\begin{table}[htbp]
\centering
\caption{Treatment-specific covariates used for DML nuisance-function estimation.}
\label{tab:ddml_covariates}
\small
\renewcommand{\arraystretch}{1.15}
\setlength{\tabcolsep}{8pt}
\begin{tabularx}{\textwidth}{>{\raggedright\arraybackslash}X c c c c}
\toprule
\textbf{Covariate}
& \textbf{ANPI}
& \textbf{FSA}
& \textbf{COI}
& \textbf{CU} \\
\midrule
\multicolumn{5}{l}{\textbf{Operational scale and demand}}\\
Capacity Utilization
&  & \cmark & \cmark &  \\
Passenger Journeys
&  & \cmark &  &  \\
Standardized Capacity km
& \cmark &  &  & \cmark \\
Network Length -- Operated
&  &  &  & \cmark \\
Car Hours -- Actual Revenue
&  &  &  & \cmark \\
Car km -- Actual Revenue Operating
&  &  &  & \cmark \\
\addlinespace
\multicolumn{5}{l}{\textbf{Fleet characteristics}}\\
Cars per Train
& \cmark & \cmark & \cmark & \cmark \\
Fleet -- Cars which are Available
& \cmark &  & \cmark & \cmark \\
Average Car Length
&  &  & \cmark & \cmark \\
Car Seating Capacity
& \cmark & \cmark & \cmark & \cmark \\
Design Space for Standing Passengers
& \cmark &  &  & \cmark \\
\addlinespace
\multicolumn{5}{l}{\textbf{Operational characteristics}}\\
Average Commercial Speed
& \cmark &  & \cmark & \cmark \\
Maintenance -- Infrastructure and Station Facilities
& \cmark & \cmark & \cmark & \cmark \\
Energy Consumption -- Total
&  &  &  & \cmark \\
Total Own Labor Costs
& \cmark & \cmark & \cmark & \cmark \\
\addlinespace
\multicolumn{5}{l}{\textbf{Economic controls}}\\
Operating Revenue
& \cmark & \cmark & \cmark & \cmark \\
PPP
& \cmark & \cmark & \cmark & \cmark \\
FX
& \cmark & \cmark & \cmark & \cmark \\
\bottomrule
\end{tabularx}
\vspace{0.3em}
\begin{minipage}{0.96\textwidth}
\footnotesize
\textit{Note:} Check marks indicate covariates included in the treatment-specific nuisance-function estimation.
\end{minipage}
\end{table}

Table~\ref{tab:ddml_baseline} reports the baseline DML estimates. ANPI and CU remain positive and statistically significant, with estimated elasticities of +0.287 and +0.391, respectively. COI retains a significant negative effect of $-0.396$, whereas FSA remains statistically insignificant.

\begin{table}[htbp]
\centering
\caption{Baseline DML estimates without Year as a nuisance covariate.}
\label{tab:ddml_baseline}
\small
\renewcommand{\arraystretch}{1.15}

\begin{tabular}{lccccc}
\toprule
\textbf{Determinant}
&
\textbf{$N$}
&
\textbf{Covariates}
&
\textbf{$\theta$}
&
\textbf{SE}
&
\textbf{$p$}
\\
\midrule

ANPI
&
522
&
11
&
+0.287
&
0.074
&
$<0.001$
\\

FSA
&
500
&
9
&
0.096
&
0.251
&
0.673
\\

COI
&
535
&
11
&
$-0.396$
&
0.188
&
$<0.05$
\\

CU
&
553
&
16
&
+0.391
&
0.098
&
$<0.001$
\\

\bottomrule
\end{tabular}

\vspace{0.3em}

\footnotesize
\textit{Notes.}
$N$ denotes the estimation sample size. Covariates denotes the number of variables included in the DML nuisance-function estimation. 

\end{table}

None of the higher-order treatment terms remains significant under DML, further indicating that the earlier nonlinear patterns are sensitive to functional-form and confounder adjustment. Because this baseline specification treats operator-year observations as independent, the analysis is subsequently extended using a fixed-effects transformation.

\subsubsection{Panel DML}

Following the Hausman test results, fixed-effects transformations are applied to all four determinants before DML estimation. Table~\ref{tab:panel_ddml_results} reports the resulting estimates.

\begin{table}[htbp]
\centering
\caption{Panel DML estimates for the four operational determinants.}
\label{tab:panel_ddml_results}
\small
\renewcommand{\arraystretch}{1.15}

\begin{tabular}{lccccc}
\toprule
\textbf{Determinant}
&
\textbf{$N$}
&
\textbf{Covariates}
&
\textbf{$\theta$}
&
\textbf{SE}
&
\textbf{$p$}
\\
\midrule

ANPI
&
522
&
12
&
+0.377
&
0.080
&
$<0.001$
\\

FSA
&
500
&
10
&
$-0.521$
&
0.251
&
$<0.05$
\\

COI
&
535
&
12
&
$-0.796$
&
0.249
&
$<0.01$
\\

CU
&
553
&
17
&
+0.489
&
0.083
&
$<0.001$
\\

\bottomrule
\end{tabular}

\vspace{0.3em}

\footnotesize
\textit{Notes.}
$N$ denotes the estimation sample size. Covariates denotes the number of variables included in the DML nuisance-function estimation. Covariate counts include Year.

\end{table}

Accounting for operator-specific heterogeneity strengthens all four estimated effects. The elasticity of ANPI increases from 0.287 to 0.377, while that of CU increases from 0.391 to 0.489 (both $p<0.001$). COI becomes more negative, changing from $-0.396$ to $-0.796$ ($p<0.01$), and FSA changes from an insignificant positive estimate to a significant negative elasticity of $-0.521$ ($p<0.05$). These changes are particularly important for the supply measures, whose effects are partly obscured when persistent differences between operators are not removed.

The coefficients imply that a 1\% increase in ANPI or CU is associated with average increases of approximately 0.38\% and 0.49\%, respectively, in the number of incidents causing delays of five minutes or longer per million train-km. By contrast, a 1\% increase in FSA or COI is associated with average reductions of approximately 0.52\% and 0.80\% in this incident rate. The larger absolute COI estimate indicates a stronger elasticity for the network-length-normalized measure within the present specification. 

The direction of these estimates is consistent with the broader empirical literature. The positive ANPI effect agrees with studies linking high passenger demand to longer dwell times, reduced recovery margins, transfer-station crowding, and poorer reliability \citep{Liu2020,Zhang2024Dynamic}. The positive CU effect is also consistent with evidence that demand--capacity mismatch and local bottlenecks increase crowding and service degradation \citep{Peftitsi2021Evaluating,Liang2024Data-driven}. Conversely, the negative FSA and COI estimates accord with findings that constrained service capacity and capacity reductions impair network performance, particularly under high demand \citep{Liu2020Effects,Zhang2021Metro}. Direct comparison of elasticity magnitudes is limited because previous studies generally use system-specific data, different reliability measures, or associational designs. The present analysis extends the international benchmarking evidence of \citet{Melo2011} by estimating within-operator elasticities across a substantially larger longitudinal panel.

\section{Conclusions}

This study estimates the causal effects of four operational determinants on metro service reliability, measured by incidents causing delays of five minutes or longer per million train-km. Compared with conventional OLS and panel regression, Double/Debiased Machine Learning (DML) allows flexible adjustment for high-dimensional and potentially nonlinear confounding relationships while reducing sensitivity to nuisance-function misspecification. The fixed-effects transformation further accounts for additive time-invariant operator heterogeneity.

The Panel DML estimates reveal a consistent pattern across three operational mechanisms. Higher Annual Network Passenger Intensity (ANPI) and Capacity Utilization (CU) are estimated to increase incident rates, whereas greater Fleet Supply Adequacy (FSA) and Car-based Operational Intensity (COI) are estimated to reduce them. Reliability therefore depends not only on passenger demand, but also on whether service provision and operating capacity expand sufficiently to accommodate it. Maintaining an appropriate balance between demand and supply is central to reliability-oriented metro planning.

Several limitations qualify the causal interpretation. First, cross-fitting was conducted at the observation level rather than by metro operator, so within-operator dependence may not be fully preserved; future work should adopt operator-level cross-fitting and cluster-robust inference. Second, Year enters the nuisance functions in demeaned form and therefore does not absorb common calendar-year shocks, which could be addressed through a two-way fixed-effects transformation. Third, residual confounding may remain because unobserved time-varying differences in management quality, engineering practices, or maintenance conditions may not be fully captured by the adjustment sets or fixed effects. Finally, reverse causality is possible if persistent reliability problems induce operators to increase service provision or fleet capacity. Addressing this endogeneity would require a credible instrumental variable, potentially combined with an IV-DML framework.

\bibliographystyle{chicago}
\bibliography{references}
\end{document}